\documentclass[CTAC,short]{cta-v2mod}

\usepackage{nicefrac}												
\usepackage{amssymb}											
\usepackage{float}													
\usepackage{lineno}												
\usepackage{enumitem}
\usepackage{multirow}											
\usepackage{subcaption}											
\usepackage{textcomp}											
\usepackage{array}													
\usepackage{url}
\usepackage{pdflscape}	
\usepackage[backgroundcolor = red]{todonotes}		
\usepackage{textgreek}											
\usepackage{color, soul}											
\usepackage{listings}												

\usepackage[numbers]{natbib}  
\usepackage{pdfpages}
\usepackage{aas_macros}
\usepackage{chngcntr}
\counterwithout{figure}{chapter}

\usepackage{longtable}

\newcolumntype{H}{>{\columncolor{ctablue}}c}
\newcolumntype{S}{>{\columncolor{lightgray}}c}

\begin{document}


\title{Origin and role of relativistic cosmic particles}
\shorttitle{Origin and role of relativistic cosmic particles}
\approver{}
\releaser{}
\docnum{}
\issue{0}
\revision{1}

\history
{
	1 & b & 2019-11-05 & Second draft &  \\ 	
	\hline 	
	1 & a & 2018-12-17 & First draft &  \\ 		
	\hline 				
}

\contributors
{
\contributors
{	
	M.\ Burtovoi & \acrshort{CISAS} \acrshort{UniPD} \acrshort{INAF-OAPD} &  Chapter~\ref{GAL}\\ 
	\hline 	
	A.\ Chernyakova & \acrshort{DCU} &  Chapter~\ref{GAL}\\ 
	\hline 	
	J.-P.\ Lenain & \acrshort{LPNHE} \acrshort{CNRS/IN2P3} & Editor \\ 	
	\hline 	
	M.\ Manganaro & \acrshort{UnivRijeka} & First draft \\ 	
	\hline 	
	P.\ Romano & \acrshort{INAF-OAB} &  Editor \\ 
	\hline 	
	F.\ Tavecchio & \acrshort{INAF-OAB} &  Chapter~\ref{BH}\\ 
	\hline 	
	L.\ Zampieri & \acrshort{INAF-OAPD} &  Chapter~\ref{GAL}\\ 
	\hline 	
	A.M.\ Brown & \acrshort{Durham} &  Early comments\\ 
	\hline 	
	S.\ Vercellone & \acrshort{INAF-OAB} & Early comments \\ 
	\hline 	
	T.\ Hassan & \acrshort{CIEMAT} &  Comments to rev b\\ 
    \hline 	
	M.\ Cerruti & \acrshort{APC} \acrshort{CNRS/IN2P3} &  Comments to rev b\\ 
	\hline 	
	C.\ Boisson  & \acrshort{LUTH} &  Comments to rev b\\ 
	\hline 	
	H.\ Sol & \acrshort{LUTH} &  Comments to rev b\\ 
	\hline 	
}}






\thispagestyle{empty}
\includepdf[pages=1,pagecommand={},scale=1, offset=0cm 0cm]{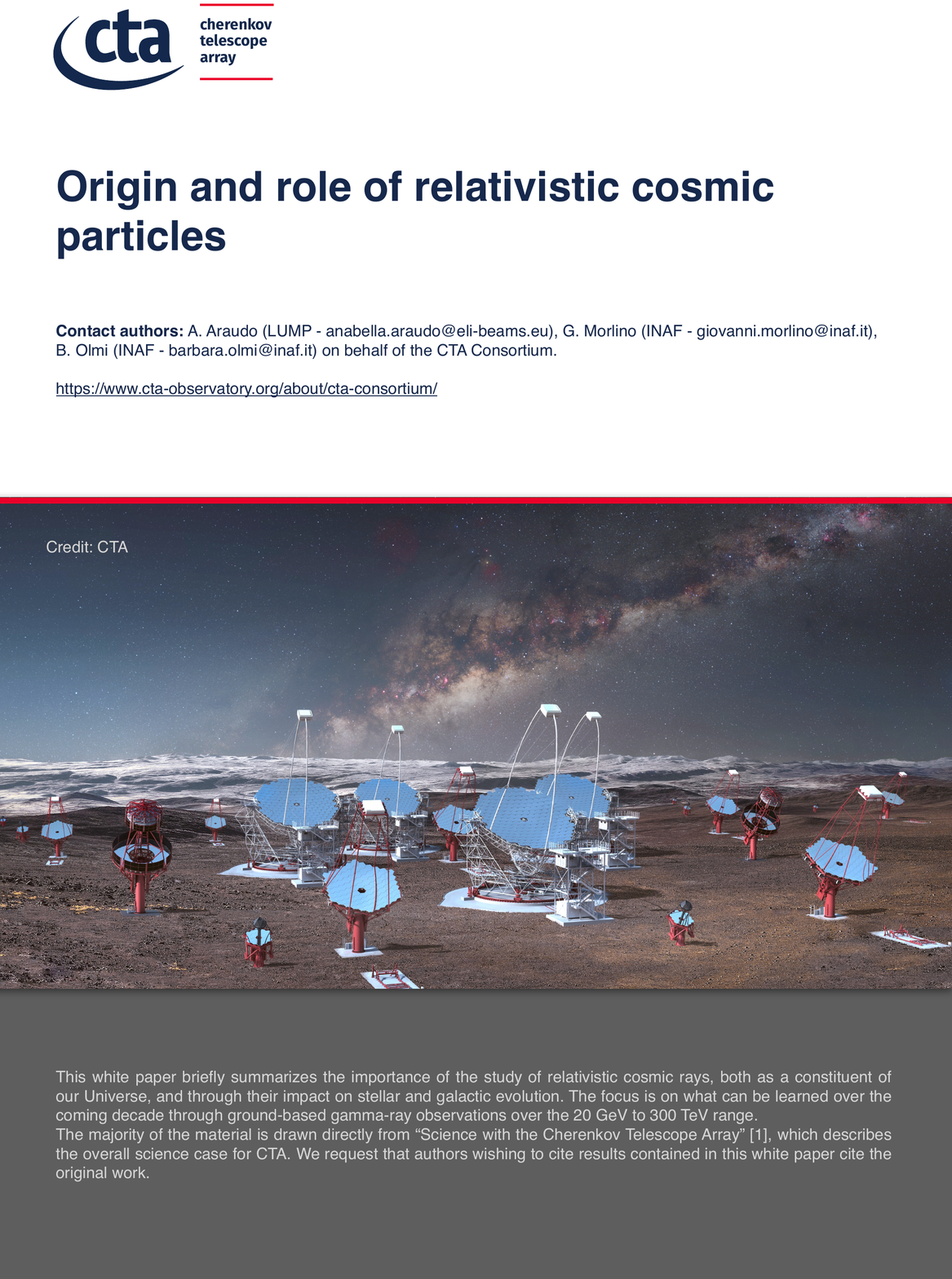}
\clearpage
\FloatBarrier \if@openright\cleardoublepage\else\clearpage\fi

%

%
\textbf{Authors}\\
%
%
F. Acero (AIM, CEA, CNRS, Université Paris-Saclay, Université de Paris, France),
I. Agudo (Instituto de Astrofísica de Andalucía-CSIC, Spain),
R. Adam (Laboratoire Leprince-Ringuet, Ecole Polytechnique, CNRS/IN2P3, 91128 Palaiseau, France),
A. Araudo (LUPM, Montpellier),
R. Alves Batista (Radboud University Nijmegen, Netherlands),
E. Amato (INAF - Osservatorio Astrofisico di Arcetri, Italy),
E.O. Anguner (Centre de Physique des Particules de Marseille, France),
L. A. Antonelli (INAF-Osservatorio Astronomico di Roma, Italy),
Y. Ascasibar (Universidad Aut\'onoma de Madrid, Spain),
C. Balazs (Monash University, Australia)
% ...
on beahlf of the CTA consortium.

%
%
\chapter*{Introduction}
Relativistic particles play a major role in a wide range of astrophysical systems, from pulsars and supernova remnants in our own Galaxy, to active galactic nuclei and clusters of galaxies. 
Within the interstellar medium of our Galaxy, these cosmic rays (CRs) are close to pressure equilibrium with interstellar gas and magnetic fields -- yet the relationship between these three components, and the overall impact on ionization and chemical evolution of the interstellar medium, the star-formation process and the evolution of galaxies, is very poorly understood.
In the coming decade, ground-based observatories have the potential to provide high-angular resolution measurements of gamma-rays resulting from the interaction of CRs with matter and photon fields, allowing to detect over-densities of CR protons and nuclei. This is in addition to gamma-ray emission associated with the energetically sub-dominant electrons that also produce the non-thermal emission seen at radio and X-ray wavelengths. 
These observations will provide insights into the mechanisms of CRs acceleration, transport, and feedback, thus making a major contribution to our deepening understanding of the processes by which galaxies and clusters of galaxies evolve.

Below we introduce the main elements of this theme, beginning with the accelerators themselves, and then moving to the wider impact of the accelerated particles. 
While we focus here on the scientific goals, and the observations required to achieve them, progress in this field over the coming decade will inevitably be dominated by the development of the Cherenkov Telescope Array (CTA) \cite{2019scta.book.....C}, a major new international observatory for GeV-TeV gamma-ray astronomy \cite{2013APh....43....3A}.
\chapter{Cosmic Accelerators}
The primary goal of gamma-ray astrophysics thus far has been to establish in which sources particle acceleration takes place and, in particular, to determine the dominant contributors to the locally measured CRs, which are 99\% protons and nuclei plus ~1\% electrons and a smaller fraction of antimatter (positrons and antiprotons). Huge progress has been made over the last decade in this area, with the combination of space-based and ground-based observations proving extremely effective in identifying the brightest Galactic accelerators, and providing strong evidence of hadron acceleration in a handful of sources. However, key questions remain unanswered: are supernova remnants (SNR) the only major contributors to the population of Galactic CRs? Where in our Galaxy are particles accelerated up to the highest (PeV$=10^{15}$ eV) energies? Which is the physical mechanism that allows particles to reach such high energies? What are the sources of high-energy cosmic electrons? How do particles escape from their accelerators?
These questions, along with the critical issue of the dominant mechanism for CRs acceleration, can be addressed over the coming decade through two main approaches:
\begin{enumerate}[label=(\roman*)]
    \item a census of particle accelerators in the Universe, achieved through sensitive Galactic and Extragalactic surveys in the 20 GeV - 300 TeV gamma-ray band, and
    \item precision gamma-ray measurements of archetypal sources, where bright nearby sources will be targeted to obtain spatially-resolved spectroscopy, or very high statistics light curves, to provide a deeper physical understanding of the processes at work in cosmic accelerators.
\end{enumerate}
A general census is required to understand the populations of CRs accelerators and the evolution and lifecycle of these source classes -- existing population studies in the TeV band are currently limited to typically only tens of objects \cite{2018A&A...612A...2H,2018A&A...612A...3H}. 
Complementary deep observations of individual sources will provide the very broad band spectra and high angular resolution images which allow to unambiguously separate leptonic from and hadronic emission, and to test acceleration to the highest energies possible for Galactic accelerators.

\begin{figure}
    \centering
    \includegraphics[scale=0.35]{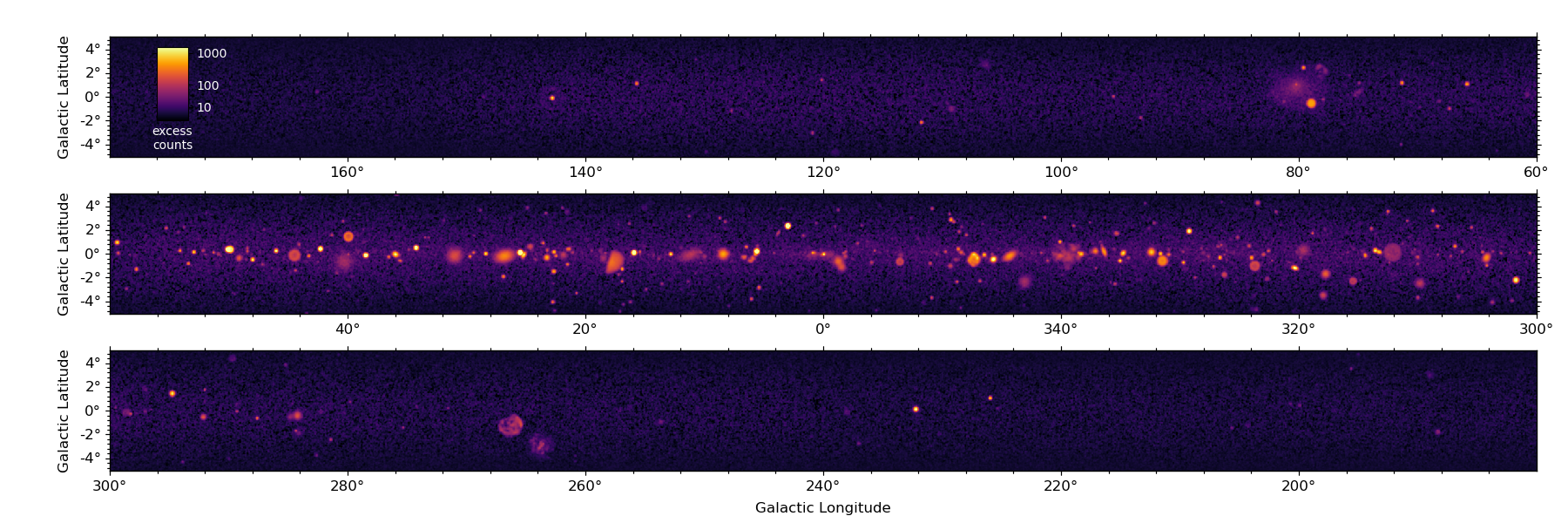}
    \caption{Simulated CTA image of the Galactic plane.}
    \label{fig:fig1}
\end{figure}

The census of particle accelerators in our Universe is best achieved by performing surveys of the sky at unprecedented sensitivity in the $>20$ GeV energy band. CTA can deliver these surveys with one to two orders of magnitude better sensitivity than existing surveys, very early in the life of the Observatory. Indeed, over much of the sky, and over much of the energy range of CTA, no survey exists, and the CTA measurements will be revolutionary. 
The observations will open up discovery space in an unbiased way and generate legacy datasets of long-lasting value. The planned survey regions include an extragalactic survey, covering $1/4$ of the extragalactic sky to a depth of 6 mCrab, and a Galactic Plane survey, with sensitivity sufficient to detect essentially the entire Galactic population of luminous ($>10^{34}$ erg/s) TeV sources, and to provide a large sample of objects up to one magnitude fainter (Fig.~\ref{fig:fig1}). 
In addition, a survey of the Large Magellanic Cloud (LMC) will provide a face-on view of an entire star-forming galaxy, resolving regions down to 20 pc in size. This will allow to map the diffuse LMC emission as well as individual objects, thus providing key information on relativistic particle transport in a galaxy different from the Milky Way.
The Galactic Plane survey will provide a complete and systematic view of the most populated regions of the Galaxy, greatly increasing our understanding of both the Galactic source populations and the diffuse emission components. It is expected to increase the catalog of known Galactic sources by a factor of 5 or more, of order 300-500 objects, primarily supernova remnants and pulsar wind nebulae (PWNe).
These last, given their long lifetime against IC radiation, are expected to be the most numerous class of sources above 100 GeV.
The identification of PWNe, especially considering that most of them will be faded away at lower energies, will represent one of the challenges of gamma-ray astrophysics in the next future. 
PWNe are well known particle accelerators and antimatter factories in the Galaxy and thus unique places to look  at to investigate particles acceleration mechanisms and escape processes, as well as definitely connected with the formation of extended TeV halos, as those revealed around a couple of evolved pulsars (Geminga and Monogem \cite{2017Sci...358..911A}).

The Galactic Plane survey will also enable comprehensive population studies of these source classes, it will reveal new and unexpected phenomena, new source classes and new forms of transient behavior, and identify candidates for the sites of acceleration of the highest energy Galactic CRs. 
These observations will address questions of how and where protons and nuclei are accelerated to PeV energies, how particles are accelerated in strong shocks or by magnetic reconnection \citep{2005A&A...441..845D} and, to some extent, how CRs impact the interstellar medium as they propagate. 
In this respect, there have been exciting recent results from the Tibet AS$\gamma$ experiment suggesting the presence of PeV hadrons both in the SNR G106.3+2.7, \citep{2021NatAs.tmp...41T} as well as in the Galactic Plane \cite{2021PhRvL.126n1101A}.
Among the most remarkable results in \cite{2021PhRvL.126n1101A} is the discovery of a diffuse gamma-ray component above 100 TeV which must be of hadronic origin. Its flux is considerable larger than predicted by models assuming space-independent diffusion. A larger flux has, therefore, to be expected also at lower energies as predicted by CR transport models with space-dependent diffusion based on Fermi-LAT data \cite{2015ApJ...815L..25G,2018PhRvD..98d3003L}.
This finding has relevant implications for the Galactic plane source survey and increases significantly the chances to probe the very high energy CRs population, hence its origin and propagation, in the inner Galactic Plane and Galactic Center region.
A deeper exposure of the inner few degrees around the Galactic Center will finally reveal the nature of the point-like central gamma-ray source, probably (but not certainly) associated with the supermassive black hole Sgr A* \citep{2019ApJ...879....6R} or with the nearby compact star clusters \citep{2019NatAs...3..561A}. 
Particle acceleration and propagation in the vicinity of the Galactic Center will be explored by studying the diffuse gamma-ray emission along the Galactic Center ridge, which extends over 1.5 degrees along the Galactic plane and is generally acknowledged to be generated by hadronic interactions \cite{2006Natur.439..695A}.

We also note that the non-thermal activity in the Galactic Centre is probably responsible for the giant Fermi Bubbles \cite{2010ApJ...724.1044S} that extend $\sim10$ kpc vertically from the Galactic Centre. CTA measurements have the potential to distinguish between different scenarios for their origin, e.g. stellar-wind driven by star formation, jet activity of the supermassive black hole or in-situ acceleration of electrons by plasma turbulence \citep{Hofmann2020}.

The Extragalactic survey will provide an unbiased population study of the local Universe ($z<0.2$) in the energy range from 100 GeV to 10 TeV. Sources in quiescent as well as flaring states will be detected, and will provide an unbiased determination of the luminosity function (log N – log S distribution) for gamma-ray emitting active galactic nuclei. This will be the first time that a such a large portion (25\%) of the sky is observed uniformly with high sensitivity at these energies, which will very likely lead to the serendipitous detection of rapid flares, and the discovery of gamma-ray emission from as yet undetected source classes such as Seyfert galaxies.

To complement these surveys, deep observations of individual sources will be performed which will have a transformational impact on our understanding over the coming decade. Objects to be targeted include supernova remnants, pulsar wind nebulae and their extended TeV haloes, gamma-ray binaries, colliding-wind binaries, massive stellar clusters, starburst galaxies and active galaxies. We emphasize that the CTA unique combination of sensitivity and high angular resolution will allow to carry out spatially-resolved spectral analyses in extended sources. This will enable the study of the diffusion and advection of energetic particles away from the pulsar in PWNe and halos and study the acceleration mechanism in different shock conditions (e.g. shock-cloud interaction) in SNRs as well as in magnetic reconnection layers in PWNe \cite{2014PhPl...21e6501C}. Along with targets selected from the existing TeV catalogs, the aforementioned surveys will provide a list of promising objects for deeper follow-up observations. In particular, the performance of CTA at the highest energies, above 20 TeV, will allow to identify and study potential Galactic PeVatrons – objects capable of accelerating CRs up to the PeV scale, in which the problematic ambiguity between leptonic and hadronic origin is almost completely resolved. There is also huge potential for the discovery of new classes of accelerators, with emission from clusters of galaxies as one of the most exciting possibilities.
\chapter{Propagation of accelerated particles and their impact on galaxies evolution}
\begin{figure}
    \centering
    \includegraphics[scale=0.42]{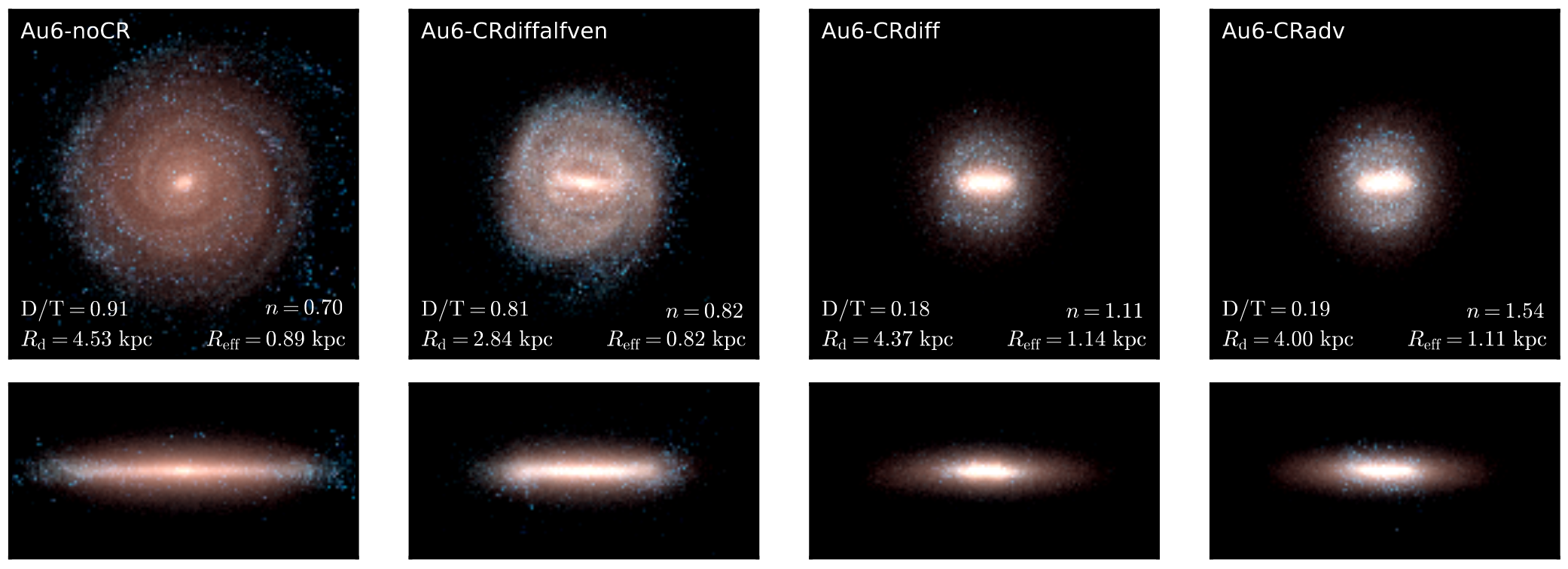}
    \caption{Face on and edge on views of the projected stellar density at $z=0$ of a  Milky Way-like galaxy, as taken from the Au6 simulation of \cite{2020MNRAS.497.1712B}. Colors represent the synthetic star luminosities: K band in red, B in green and U in blue. From left to right each panel shows: the AURIGA \cite{2017MNRAS.467..179G,2016MNRAS.462.2603P,2017MNRAS.465.4500P} simulation without CRs; the AURIGA simulation including CRs advection and anisotropic diffusion, with Alfvén cooling/heating; the simulation with  CRs advection and anisotropic diffusion; the simulation with pure CRs advection.}
    \label{fig:fig2}
\end{figure}

Beyond the question of how and where particles are accelerated in the Universe, is the question of what role these particles play in the evolution of their host objects and how they are transported out to large distances. On the scale of clusters of galaxies, CRs with TeV– PeV energies are thought to be confined over a Hubble time \citep{1996SSRv...75..279V}. 
On smaller scales, they typically escape from their acceleration sites and may impact their environments in a number of ways:
(i) as a dynamical constituent of the medium; (ii) through generation/amplification of magnetic fields, and (iii) through ionization and subsequent impact on the chemical evolution of (for example) dense cloud cores.
All of these effects are relevant for the interstellar medium of our own Galaxy and are likely to be important in star-forming systems on all scales. The first aspect is also likely to be important for the process of AGN feedback on the host galaxy cluster and growth of massive galaxies. Finally, point (ii) could be relevant also for the generation of magnetic fields in the intra-cluster medium by means of escaping CRs.
As a remarkable example of CRs feedback, in Fig.~\ref{fig:fig2} we report results from magneto-hydrodynamic simulations showing the impact that different models of CRs propagation can have on the evolution of a galaxy like the Milky Way \cite{2020MNRAS.497.1712B}. In particular, CRs affect the accretion of gas onto galaxies, modifying some structural properties such as disc sizes, densities and temperatures of the circum-galactic medium.

CTA will map extended emission around many gamma-ray sources and look for energy dependent morphology associated with diffusion (in the case of hadrons) or cooling (in the case of electrons). As the energy dependence is expected to be opposite in the two cases, such mapping provides another means, in addition to spectral studies, to separate the emission from these two populations. It is CTA’s unprecedented (in the gamma-ray domain) angular resolution, energy resolution and background rejection power that will make this possible. These characteristics will maintain CTA extremely competitive with respect to other gamma-ray instruments (as LHAASO and Tibet AS$\gamma$), with unsurpassed performances at any energy above 10 TeV.
Among the targets are star-forming systems -- both star-forming regions within our Galaxy and the LMC, as well as nearby spiral and starburst galaxies. Star forming regions are, in fact, direct competitors of SNRs in the production of relativistic particles, thanks to the powerful winds of massive stars able to produce shocks and to release a total kinetic energy comparable to the one released by SNRs \cite{2019NatAs...3..561A,2021NatAs.tmp...57A}. Observations with CTA will probe the relationship between star formation and particle acceleration in these systems. 
On the scale of individual galaxies, the LMC provides a unique target hosting a number of extraordinary objects, including 30 Doradus - the most active star-forming region in the local group of galaxies, the super star cluster R136, supernova SN1987A and the 30 Dor C superbubble. It is one of the nearest star-forming galaxies, with one-tenth of the star formation rate of the Milky Way distributed in only two percent of its volume. Since it is observed face-on, at high Galactic latitudes, observations of the LMC will provide a well-resolved global view of a star-forming galaxy at very high energies, allowing to study the transport of CRs from their release into the interstellar medium to their escape from the system. Nearby spiral, starburst and ultra-luminous infrared galaxies (ULIRGs) provide additional galaxy-scale objects for CRs studies. The enhanced rate of star formation in these systems is expected to lead to intense CRs production through associated supernovae - in an extreme object such as the nearby ULIRG Arp 220, a supernova explosion is expected to occur once every 6 months \cite{1998ApJ...493L..17S}, as compared to roughly once per century in our own Galaxy. This is coupled with an enhanced density of low energy photons and interstellar material to act as targets for gamma-ray production mechanisms from both leptonic and hadronic CRs.

The identification of the sources responsible for the production of UHECR with $E>10^{17}$ eV, is also a very important task. Unfortunately, if the sources are located at cosmological distances ($z>0.1$), the high energy radiation produced by those particles inside the accelerator is absorbed either by the photon-field of the source itself or by the extragalactic background light (EBL). However, a hadronic beam (HB) travelling towards the Earth produces electromagnetic cascades in the intergalactic space. Because of the reduced distance, high-energy gamma-rays produced by the cascades experience a less severe absorption by the interaction with the EBL and can reach the Earth \cite{2010APh....33...81E}. Hence, a distinctive prediction of this model is that the observed gamma-ray spectrum extends at energies much higher than those allowed by the conventional propagation through the EBL. For sources located at low red-shift ($z<0.3$), the spectra should be characterized by a hard tail above 10 TeV, whose detection is considered the smoking gun of this model \cite{2012ApJ...749...63M}. CTA will be able to test this scenario thanks to its high sensitivity above 10 TeV.

On the largest scales, the most massive gravitationally bound systems in the Universe - galaxy clusters – are expected to be reservoirs of CRs accelerated both by structure formation processes, and by their constituent galaxies and active galactic nuclei. Cosmic ray protons in the intra-cluster medium should accumulate over cosmological timescales \cite{1996SSRv...75..279V,1997ApJ...487..529B}, leading to subsequent gamma-ray emission. To date, no Galaxy clusters have been unambiguously detected in gamma-rays with the possible exception of the Coma cluster observed by Fermi-LAT \cite{2021A&A...648A..60A}. However, based on theoretical studies and hydrodynamical simulations, the Perseus cluster should be the brightest target, detectable with CTA. Deep observations of Perseus with CTA will determine the CRs proton content of clusters and measure its dynamical impact on the cluster environment.
%


{\em Acknowledgements.} 
We gratefully acknowledge financial support from the agencies and organizations listed here: \href{http://www.cta-observatory.org/consortium_acknowledgments}{http://www.cta-observatory.org/consortium\_acknowledgments}.

\bibliographystyle{apsrev}
\bibliography{biblio}


\glsaddall
\setglossarystyle{list}
\printglossaries

\end{document}